\documentstyle[11pt,newpasp,twoside,epsf]{article}
\def\edcomment#1{\iffalse\marginpar{\raggedright\sl#1\/}\else\relax\fi}
\reversemarginpar

\begin{document}
\title{3-D General Relativistic MHD Simulations of Generating Jets}
 \author{K.-I. Nishikawa}
\affil{Rutgers University, Department of Physics and Astronomy,
Piscataway, NJ 08854-8019, USA}
\author{S. Koide}         
\affil{Toyama University, Faculty of Engineering, 3190 Gofuku, Toyama 930-8555,
   Japan}
\author{K. Shibata}       
\affil{Kyoto University, Kwasan Observatory, Yamashina, Kyoto 607-8471, Japan}
\author{T. Kudoh}         
\affil{National Astronomical Observatory, Mitaka, Tokyo 181-8588, Japan}
\author{H. Sol}         
\affil{DARC, Observatoire de Paris-Meudon, 92195 Meudon Cedox, France}

\begin{abstract}
   We have performed a first fully 3-D GRMHD simulation with Schwarzschild black
hole with a free falling corona. The initial simulation results show that a jet
is created as in previous axisymmetric simulations. However, the time to 
generate the jet is longer than in the 2-D simulations. We expect that due to
the additional azimuthal dimension the dynamics of jet formation can be 
modified.
\end{abstract}

\section{Introduction}

   Relativistic jets have been observed in active galactic nuclei (AGNs) and
microquasars in our Galaxy, 
and it is believed that they originate
from the regions near accreting black holes. To investigate the
dynamics of accretion disks and the associated jet formation, we use our 
newly developed full 3-D GRMHD code. 

   Recently,  Koide, Shibata, \& Kudoh (1999) have investigated the dynamics
of an accretion disk initially threaded by a uniform poloidal magnetic field
in a non-rotating corona (either in a state of steady fall 
or in hydrostatic equilibrium) around 
a non-rotating black hole. The numerical results
show that the disk loses angular momentum by magnetic braking, then 
falls into 
the black hole.  The infalling motion of the disk, 
which is faster than in 
the non-relativistic case because of the general-relativistic 
effect below $3 r_{\rm S}$ ($r_{\rm S}$ is the Schwarzschild radius),
is strongly decelerated at the shock formed by the centrifugal force
around $r = 2 r_{\rm S}$ by the rotation
of the disk. Plasmas near the shock are accelerated by the ${\bf J}
\times {\bf B}$ force, which forms bipolar relativistic jets. Inside this
{\it magnetically driven jet}, the gradient of gas pressure 
also generates a jet
over the shock region ({\it gas-pressure-driven jet}). 
This {\it two-layered jet structure} is formed both in a hydrostatic corona
and in a steady-state falling corona. Koide et al.~(2000) have also developed
a new GRMHD code in Kerr geometry and have found that, 
with a rapidly rotating ($a = 0.95$) black-hole magnetosphere,  
the maximum velocity of the jet is 
0.9 c and its terminal velocity 0.85 c. 
All of the previous 2-D GRMHD simulations described here were made 
assuming axisymmetry with respect
to the $z$-axis and mirror symmetry with respect to the plane $z = 0$;
the axisymmetric assumption suppressed the azimuthal instabilities. 

\section{3-D GRMHD Simulations: Equations and Numerical Techniques}

Our basic equations are those of Maxwell for the fields  
and a set of general-relativistic equations representing the plasma, namely 
the equations of conservation of mass, momentum, and energy for a 
single-component conducting fluid 
(Weinberg 1972; Thorne et al. 1986).  In making 
the simulations, we use these equations with the 3+1 formalism 
(for details, see Koide, Shibata, \& Kudoh 1999).  

\section{ Preliminary 3-D GRMHD simulations with a Schwarzschild 
    black hole}

In order to investigate how accretion disks near
black holes evolve under the influence of accretion instabilities
such as the magnetorotational instability, the use of a fully 3-D GRMHD is
essential.                                                                            

\subsection{Initial and boundary conditions}

In the assumed initial state, the simulation region is divided into two parts: 
a background corona around a black hole, and an accretion disk (Fig. 1a and 
1b). The coronal plasma is set in a state of transonic free-fall flow, 
as in the case of the transonic flows with $\Gamma = 5/3$ and $H=1.3$; here 
the sonic point is located at $r=1.6 r_{\rm S}$.
The Keplerian disk in the corona is set in the following way. 
The disk region is located at
$ r > r_{\rm D} \equiv 3r_{\rm S},
| {\rm cos} \theta | < \delta =1/8$.
Here 
the density is 100 times that of the 
background corona
(Fig. 1a), while the orbital velocity is relativistic 
and purely azimuthal: 
$v_\phi = v_{\rm K} \equiv c/[2(r/r_{\rm S} -1)]^{1/2}$.
(Note that this equation reduces to the Newtonian
Keplerian velocity $v_\phi =\root \of {GM/r}$ in 
the non-relativistic
limit $r_{\rm S}/r \ll 1$). 
The pressure of both 
the corona and the disk are assumed equal 
to that of the transonic solution.
The initial conditions for the entire 
plasma around the black hole are: 
\begin{equation}
\rho = \rho _{\rm ffc} +\rho _{\rm dis}
\end{equation}

\begin{equation}
\rho _{\rm dis} = \left \{  \begin{array}{cc}
100 \rho _{\rm ffc} &
\verb!   ! ( r > r_{\rm D} \verb!  ! {\rm and} \verb!  !
|{\rm cot} \theta| < \delta )\\ 0                  &
\verb!   ! ( r \leq r_{\rm D} \verb!  ! {\rm or} \verb!  !
|{\rm cot} \theta| \geq \delta )
\end{array} \right .
\end{equation}

\begin{equation}
(v_r, v_\theta , v_\phi) = \left \{  \begin{array}{cc}
(0, 0, v_{\rm K}) &
\verb!   ! ( r > r_{\rm D} \verb!  ! {\rm and} \verb!  !
|{\rm cot} \theta| < \delta )\\ (-v_{\rm ffc}, 0, 0)        &
\verb!   ! ( r \leq r_{\rm D} \verb!  ! {\rm or} \verb!  !
|{\rm cot} \theta| \geq \delta )
\end{array} \right .
\end{equation}
where we set $\delta = 0.125$; the 
smoothing length is $0.3 r_{\rm S}$.

In addition, there is a magnetic field crossing 
the accretion disk perpendicularly.  
We set it to the Wald solution (Wald 1974), 
which represents the uniform magnetic field 
around a Kerr black hole:
$B_r = B_0 {\rm cos} \theta$,
$B_{\theta } = - \alpha B_0 {\rm sin} \theta $ (where
$\alpha$ is the lapse function, $\alpha = (1 - r_{\rm S}/r)^{1/2}$).
At the inner edge of the accretion disk,
the proper Alfv\'{e}n velocity is $v_{\rm A} = 0.015c$
in a typical case with $B_0 =0.3 \, \root \of {\rho _0c^2}$,
where the 
Alfv\'{e}n velocity in the fiducial observer 
\begin{equation}
v_{\rm A} \equiv B/{\root \of {\rho + [\Gamma p/(\Gamma -1)+B^2]/c^2}}.
\end{equation}
The plasma beta of the corona at $r=3r_{\rm S}$
is $\beta \equiv p/B^2=1.40$.
The simulation is performed in the region
$1.1 r_{\rm S} \leq r \leq 20 r_{\rm S}$, $0 \leq \theta \leq
\pi$, $0 \leq \phi \leq 2\pi$ with $100 \times 120 \times 60$ meshes.
The effective linear 
mesh widths at $r=1.1r_{\rm S}$ and at 
$r=20r_{\rm S}$
are $5.38 \times 10^{-3} r_{\rm S}$ and $0.97 r_{\rm S}$,
respectively, while the angular spacings along the polar and azimuthal
directions are $5.2 \times 10^{-2}$ rad. 
A radiative boundary condition is imposed at $r=1.1 r_{\rm S}$ and at 
$r=20 r_{\rm S}$:
\begin{equation}
u_0^{n+1} = u_0^n + u_1^{n+1} - u_1^n ,
\end{equation}
where the 
superscripts $n+1$ and $n$ denote the time steps and
the subscripts 0 and 1 refer to the boundary and to
its neighbor meshes, respectively.
The computations were made on an ORIGIN 2000 computer
with 0.898 GB internal memory, and they used
about 47 hours of CPU time for 10000 time steps
with $100 \times 120 \times 60$ meshes.                                               

\subsection{Simulation results}

Figure 1 shows the evolution of 3-D simulation
performed in the region
$1.1 r_{\rm S} \le r \le 20 r_{\rm S}$,  $0 \le \theta \le \pi$,
and $0 \le \phi \le 2\pi$ with $100 \times 120 \times 60$ meshes.
The parameters used in this simulation are the same as those of the
axisymmetric simulations shown in Fig. 6 of Koide, Shibata, \& Kudoh (1999).
In this figure, the colored shading  
shows the proper mass density ((a) and (c)) and the 
pressure ((b) and (d)) on logarithmic scales; the 
vector plots show the magnetic field ((a) and (c)) and flow velocity ((b) and 
(d)).

The black circle represents the black hole. Figures 1a and 1b present 
the initial conditions, which are the same as in the 2-D simulations (Koide,
Shibata, \& Kudoh 1999). At  $t= 39.2 \tau_{\rm S}$ comparing with 
Fig. 6c ($t= 40.0 \tau_{\rm S}$) in Koide et al. (1999) the jet is less
generated. At $t= 60.0 \tau_{\rm S}$ the jet is generated as 2-D simulation
at the earlier time ($t= 40.0 \tau_{\rm S}$). At $t= 73.9 \tau _{\rm S}$, 
the jet is clearly created around $r = 4.5 r_{\rm S}$, which is shown by the 
enhanced density (Fig. 1c) and pressure (Fig. 1d). As in the 2-D simulations
the jet is generated in a hollowed cylindrical form. The magnetic field 
is twisted by the accretion disk and pinched, which increases the magnetic
field pressure and generates the jet near the black hole. The delay
of jet formation seems to be due to reduction of shock formation at 
$r = 2 r_{\rm S}$ caused by the additional freedom in the azimuthal
direction. Further investigation will be reported elsewhere.

\newpage
\begin{figure}
\plotfiddle{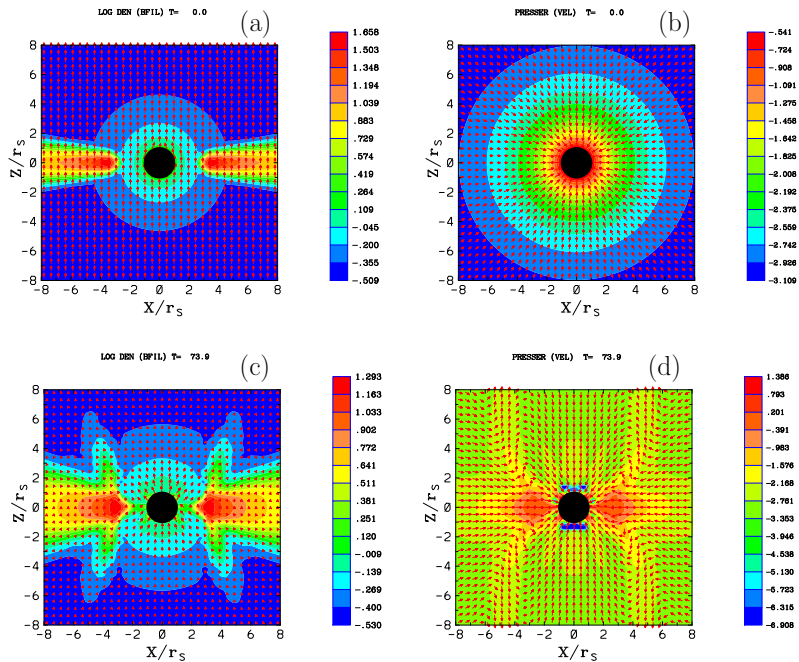}{9.0truecm}{0.}{60}{60}{-200.}{-160.}
\caption{For the fully 3-D simulation,these panels present the time
evolution of the proper mass density with the magnetic field ($B_{x}, B_{z}$)
((a) and (c)) and the proper pressure with the flow velocity 
($v_{x}, v_{z}$) ((b) and (d)) in a transonic free-fall (steady-state 
falling) corona with an initially
uniform magnetic field, at $t=0.0\tau _{\rm S}$ ((a) and (b)),
and $t=73.9 \tau_{\rm S}$ ((c) and (d)). The jet is formed around $r = 4.5
r_{\rm S}$ as in the 2-D simulation.} 
\end{figure}  
\section{Discussion}
  This simulation result is initial and we will perform more simulations
and investigate effects of the third dimension. Further results will be 
reported elsewhere. K.N. is partially supported by NSF ATM 9730230,    
ATM-9870072, ATM-0100997, and INT-9981508. The simulations have been performed 
on ORIGIN 2000 at NCSA which is supported by NSF.                                 


\begin{references}
\reference
Koide, S., Shibata, K., \& Kudoh, T. 1999, \apj, 522, 727
\reference 
Koide, S., Meier, D.~L., Shibata, K., \& Kudoh, T.  2000, ApJ, 536, 668 
\reference 
Thorne, K.~S., Price, R.~H., \& Macdonald, D.~A. 1986, Black Holes: The 
Membrane Paradigm (New Haven: Yale Univ. Press)
\reference 
Wald, R.~M. 1974, \prd, 10, 1680
\reference 
Weinberg, S. 1972, Gravitation and Cosmology (New York: Wiley)
\end{references}
\end{document}